\documentclass[twocolumn,showpacs,aps,prd,floatfix]{revtex4}

\usepackage{amsmath}
\usepackage{amsthm}
\usepackage{amssymb}
\usepackage{graphicx}
\usepackage{dcolumn}
\usepackage{ifthen}

\newboolean{physMode}
\setboolean{physMode} {true}

\newboolean{jourMode}
\setboolean{jourMode} {true}

\newboolean{confFlag}
\setboolean{confFlag} {false}

\newboolean{authFlag}
\setboolean{authFlag} {true}

\newcommand {\BADnumber} {\jourNUM}
\newcommand {\BADdate}   {\today}
\newcommand {\BADver}    {15.0}

\newcommand{\BABARPubYear}     {09}
\newcommand{\BABARPubNumber}  {003}
\newcommand{\SLACPubNumber} {13592}

\input{pubboard/babarsym.tex}

\long\def\inst#1{\par\nobreak\kern 4pt\nobreak
    {\it #1}\par\vskip 10pt plus 3pt minus 3pt}

\newcommand{\jourNUM} {2078}

\newcommand{\belle} {Belle}

\newcommand{\Cherenkov} {Cherenkov}

\newcommand{\geantfour} {GEANT4}

\newcommand{\ROOT} {ROOT}

\newcommand{\deltaE} {\mbox{$\Delta E$}}

\newcommand{\pdf} {\mbox{\rm PDF}}

\newcommand{\like} {\ensuremath{\cal{L}}}

\newcommand{\mlp}   {\ensuremath{m(\bar{\Lambda} p)}}

\newcommand{\mlbp}   {\ensuremath{m(\bar{\Lambda} p)}}

\newcommand{\splot} {\ensuremath{\mbox{}_s\cal{P}}lot}

\newcommand{\blppi} {\ensuremath{B^0\rightarrow \bar{\Lambda} p \pi^-}}

\newcommand{\bbarlppi} {\ensuremath{\bar{B^0}\rightarrow \Lambda \bar{p} \pi^+}}

\newcommand{\ELambdaB} {\ensuremath{E^*_{\Lambda}}}

\newcommand{\ELambdabar} {\ensuremath{E^*_{\bar{\Lambda}}}}

\newcommand{\ThetaH} {\ensuremath{\theta_{\rm H}}}

\newcommand{\CosThetaH} {\ensuremath{\cos\theta_{\rm H}}}

\newcommand{\Chebyshev} {Chebyshev}

\newcommand{\BelleOnPeakLumi} {414}
\newcommand{\BelleBlppiMeasurement} {\mbox{$[
3.23^{+0.33}_{-0.29} ({\rm stat.})\pm0.29 ({\rm syst.})]
\times 10^{-6}$}}
\newcommand{\TotalOnPeakLumi} {426}
\newcommand{\TotalBPairs} {467}
\newcommand{\TotalBPairsSysErr} {1.1}
\newcommand{\TotalBPairsPerc} {$\TotalBPairsSysErr\%$}

\newcommand{\DeltaEWinCut} {27}
\newcommand{\MesLowWinCut} {5.274}
\newcommand{\PLHVeryLoosePKLikeRatioMin} {0.33}
\newcommand{\PLHVeryLoosePPiLikeRatioMin} {1}

\newcommand{\GoodTracksLoosePtMin} {50}

\newcommand{\PBbkgKMSuperLooseEffNoErr} {$17.8\%$}
\newcommand{\PBmcsigCumKMSuperLooseEffNoErr} {$96.4\%$}

\newcommand{\PiBbkgKMSuperLooseEffNoErr} {$66.8\%$}
\newcommand{\PiBmcsigCumKMSuperLooseEffNoErr} {$98.8\%$}

\newcommand{\FishermcsigEffNoErr} {$72\%$}
\newcommand{\FisherbkgEffNoErr} {$8\%$}

\newcommand{\VtxCutmcsigEffNoErr} {$94.4\%$}
\newcommand{\VtxCutbkgEffNoErr} {$58.2\%$}

\newcommand{\VtxProbCutL} {\mbox{$10^{-6}$}}

\newcommand{\LambdaInvMassCutL} {1.111}
\newcommand{\LambdaInvMassCutU} {1.121}
\newcommand{\LambdaFlightSigCutL} {20}

\newcommand{\LmFlCutdataSideRejNoErr} {$42\%$}
\newcommand{\LmFlCutmcsigSideEffNoErr} {$90\%$}

\newcommand{\DalitzBoxN} {$20\times20$}

\newcommand{\SCFExpectedFractionNoErr} {$0.006$}

\newcommand{\LambdaCInvMassCut} {20}
\newcommand{\LambdaCInvMassCutSigma} {5}
\newcommand{\DeltaEFitRegion} {100}
\newcommand{\MesFitRegion} {5.2}
\newcommand{\MesFitRegionLow} {5.20}
\newcommand{\MesFitRegionUp} {5.29}

\newcommand{\FDCutSystematicNoPerc} {1.0}
\newcommand{\VtxCutSystematicNoPerc} {1.0}
\newcommand{\BZeroOverBChargedDefault} {$50\%$}
\newcommand{\BZeroOverBChargedFraction} {$(48.4\pm 0.6)\%$}

\newcommand{\TotalBPairsSystematicNoPerc}{\TotalBPairsSysErr}

\newcommand{\TotalTrackSystematicNoPerc} {2.4} 
\newcommand{\TotalTrackSystematic} {\TotalTrackSystematicNoPerc\%}
\newcommand{\PIDSystematicNoPerc} {1.4}
\newcommand{\PIDSystematic} {\PIDSystematicNoPerc\%}
\newcommand{\DalitzStatSystematicNoPerc} {0.4}
\newcommand{\DalitzStatSystematic} {\DalitzStatSystematicNoPerc\%}
\newcommand{\LambdaFlightSystematicNoPerc}{2.8}

\newcommand{\LambdaMassSystematicNoPerc} {2.4}

\newcommand{\SignalPDFSystematic} {3.2\%}
\newcommand{\BackgroundPDFSystematic} {2.2\%}
\newcommand{\LikelihoodParametersNoPerc} {3.9}
\newcommand{\LikelihoodParameters} {\LikelihoodParametersNoPerc\%}
\newcommand{\DeltaESystematicNoPerc} {1.7}
\newcommand{\DeltaESystematic} {\DeltaESystematicNoPerc\%}
\newcommand{\SCFSystematicNoPerc} {0.8}

\newcommand{\LikelihoodBiasNoPerc} {0.6}
\newcommand{\LikelihoodBias} {\LikelihoodBiasNoPerc\%}
\newcommand{\LambdaCPiResBackSystematic} {0.2\%}
\newcommand{\BLambdaPPiVetoSystematicNoPerc}{0.5}
\newcommand{\BLambdaPPiVetoSystematic} {\BLambdaPPiVetoSystematicNoPerc\%}

\newcommand{\LambdaCVetoSystematicNoPerc}{\BLambdaPPiVetoSystematicNoPerc}
\newcommand{\LambdaPPiBFSystematicNoPerc}{0.8}

\newcommand{\BFractionSystematicNoPerc} {3.2}
\newcommand{\BFractionSystematic} {\BFractionSystematicNoPerc\%}
\newcommand{\MesSystematic} {1.9\%}
\newcommand{\DESystematic} {1.3\%}
\newcommand{\TotalSystematicNoPerc} {7.4}

\newcommand{\DeltaEScaleDiffPerc} {5\%}

\newcommand{\LambdaPPiBF} {\mbox{$0.639\pm0.005$}}

\newcommand{\LambdaAlpha} {\mbox{$0.642\pm0.013$}}

\newcommand{\TotLikePolarizSystematicA} {0.004}
\newcommand{\TotLikePolarizSystematicB} {0.03}
\newcommand{\TotLikePolarizSystematicC} {0.03}

\newcommand{\MaxModulationPolarizSystematicA} {0.05}
\newcommand{\MaxModulationPolarizSystematicB} {0.07}
\newcommand{\MaxModulationPolarizSystematicC} {0.04}

\newcommand{\DataSampleSelectedEvents} {6360}

\newcommand{\LikeNSFitResult} {$183.3^{+19.2}_{-18.5}$}
\newcommand{\LikeNBFitResult} {$6176\pm80$}
\newcommand{\LikeMuDEFitResult} {$-2.65\pm 1.84$} 
\newcommand{\LikeCOneDEFitResult} {$-3.5\pm 0.4$}
\newcommand{\LikeMuMesFitResult} {$5.2797\pm 0.0003$} 
\newcommand{\LikeCArgusMesFitResult} {$-14.6\pm 1.45$}

\newcommand{\SPlotEffCorrectedYield} {$916\pm 92$}

\newcommand{\FinalBFVal} {3.07} 
\newcommand{\FinalBFErrStat} {0.31} 
\newcommand{\FinalBFErrSyst} {0.23} 

\newcommand{\FinalBFCombined} {\mbox{$\left[
\FinalBFVal\pm\FinalBFErrStat ({\rm stat.})\pm\FinalBFErrSyst ({\rm syst.})\right]
\times 10^{-6}$}}

\newcommand{\FinalAcpVal} {-0.10}
\newcommand{\FinalAcpErrStat} {0.10}
\newcommand{\FinalAcpErrSyst} {0.02}

\newcommand{\FinalAcpCombined} {\mbox{$
\FinalAcpVal\pm\FinalAcpErrStat ({\rm stat.})
\pm\FinalAcpErrSyst ({\rm syst.})$}}

\newcommand{\ElbMaxLikeBinNumberLett} {three}
\newcommand{\ElbMaxLikeRangeLow} {1.1}
\newcommand{\ElbMaxLikeRangeHigh} {2.4}

\newcommand{\ElbMaxLikeBinZeroLow} {\ElbMaxLikeRangeLow 0}
\newcommand{\ElbMaxLikeBinOneLow} {1.53}
\newcommand{\ElbMaxLikeBinTwoLow} {1.80}
\newcommand{\ElbMaxLikeBinTwoHigh} {\ElbMaxLikeRangeHigh 0}

\newcommand{\HeliBkgChebyDegree} {fourth}
\newcommand{\HeliBkgChebyCoeffN} {four}

\newcommand{\ElbCosLHeliEffBoxN} {\ensuremath{20\times20}}

\newcommand{\HeliMesSideband} {5.27}

\newcommand{\DataSampleElbCutSelectedEvents} {3994}

\newcommand{\HeliBinZeroNB} {$519 \pm 23$}
\newcommand{\HeliBinZeroNS} {$63 \pm 9$}

\newcommand{\HeliBinOneNB} {$643 \pm 26$}
\newcommand{\HeliBinOneNS} {$51 \pm 9$}

\newcommand{\HeliBinTwoNB} {$2663 \pm 52$}
\newcommand{\HeliBinTwoNS} {$55 \pm 11$}

\newcommand{\HeliBinZeroPolariz} {$-0.08^{+0.47}_{-0.40}\pm0.09$}

\newcommand{\HeliBinOnePolariz} {$+0.64^{+0.73}_{-0.65}\pm0.12$}

\newcommand{\HeliBinTwoPolariz} {$+0.97^{+0.62}_{-0.62}\pm0.08$}

\newcommand{\HeliTBinZeroPolariz} {$+0.25^{+0.53}_{-0.58}\pm0.09$}

\newcommand{\HeliTBinOnePolariz} {$+0.56^{+0.42}_{-0.48}\pm0.12$}

\newcommand{\HeliTBinTwoPolariz} {$+0.05^{+0.61}_{-0.60}\pm0.08$}

\newcommand{\HeliNBinZeroPolariz} {$-0.64^{+0.34}_{-0.33}\pm0.09$}

\newcommand{\HeliNBinOnePolariz} {$-0.78^{+0.39}_{-0.36}\pm0.12$}

\newcommand{\HeliNBinTwoPolariz} {$+0.26^{+0.53}_{-0.53}\pm0.08$}

\begin{document}

\ifthenelse{\boolean{authFlag}}{}{
\begin{flushleft}
  \babar-PUB-\BABARPubYear/\BABARPubNumber \\
  SLAC-PUB-\SLACPubNumber \\
\end{flushleft}
}
\smallskip

\title{
\large \bf \boldmath
 Measurement of the Branching Fraction and $\bar{\Lambda}$ Polarization in
$B^0\rightarrow \bar{\Lambda} p \pi^-$}
 \smallskip

\ifthenelse{\boolean{authFlag}}{}{
\begin{flushright}
BAD~\BADnumber, Version~\BADver\\
 \BADdate
\end{flushright}
}

%
%
\ifthenelse{\boolean{authFlag}}{
%
\author{B.~Aubert}
\author{Y.~Karyotakis}
\author{J.~P.~Lees}
\author{V.~Poireau}
\author{E.~Prencipe}
\author{X.~Prudent}
\author{V.~Tisserand}
\affiliation{Laboratoire d'Annecy-le-Vieux de Physique des Particules (LAPP), Universit\'e de Savoie, CNRS/IN2P3,  F-74941 Annecy-Le-Vieux, France}
\author{J.~Garra~Tico}
\author{E.~Grauges}
\affiliation{Universitat de Barcelona, Facultat de Fisica, Departament ECM, E-08028 Barcelona, Spain }
\author{M.~Martinelli$^{ab}$}
\author{A.~Palano$^{ab}$ }
\author{M.~Pappagallo$^{ab}$ }
\affiliation{INFN Sezione di Bari$^{a}$; Dipartimento di Fisica, Universit\`a di Bari$^{b}$, I-70126 Bari, Italy }
% ...
\collaboration{The \babar\ Collaboration}
\noaffiliation

}{}

%
%
\begin{abstract}
%
%
%
%
%

 We present a measurement of the $B^0\rightarrow \bar{\Lambda} p \pi^-$
branching fraction performed using the \babar\ detector at the \pep2\
asymmetric \epem\ collider. Based on a sample of $\TotalBPairs\times 10^6$
$B\bar{B}$ pairs we measure ${\cal{B}}(B^0\rightarrow \bar{\Lambda} p
\pi^-) = \FinalBFCombined.$ The measured differential spectrum as
a function of the dibaryon invariant mass \mlbp\ 
shows a near-threshold enhancement similar to that observed in other
baryonic B decays. We study the $\bar{\Lambda}$ polarization as a
function of $\bar{\Lambda}$ energy in the $B^0$ rest frame ($\ELambdabar$) and
compare it with theoretical expectations of fully longitudinally right-polarized
$\bar{\Lambda}$ at large $\ELambdabar$.
\end{abstract}

\pacs{13.25.Hw; 13.60.Rj}

\maketitle

%
%
%
%
%
%
%

\ifthenelse{\boolean{confFlag}}{
\section{INTRODUCTION}
}{
\section{Introduction}
}

\label{sec:Introduction}

Observations of charmless three-body baryonic $B$ decays have been
reported recently by both the \belle\ and \babar\ collaborations
\cite{ref:Belle_Blm_Observ,ref:Belle_Blm_Update,ref:BaBar_Bppk_Study}. A
common feature of these decay modes is the peaking of the
baryon-antibaryon mass spectrum near threshold. This feature has
stimulated considerable interest among theorists as a key element in the
explanation of the unexpectedly high branching fractions for these decays
\cite{ref:Hou_Soni_PRL86,ref:Chua_Hou_EurC29}.

In the standard model, the \blppi\ decay proceeds through tree level $b\rightarrow
u$ and penguin $b\rightarrow s$ amplitudes. It is of interest to study the
structure of the decay amplitude in the Dalitz plane to test
theoretical expectations. 
The weak decay $\bar{\Lambda}\rightarrow \bar{p} \pi^+$ is spin self-analyzing.
Since the $\bar{s}$ quark carries the $\bar{\Lambda}$ spin, the V-A 
transition $b\rightarrow s$  leads to the expectation that the $\bar{\Lambda}$
is fully longitudinally right-polarized at large $\bar{\Lambda}$ energy in the 
$B^0$ rest frame \cite{ref:Suzuki_JPG29}. This channel may also be used to
search for direct \CP\ violation.

\ifthenelse{\boolean{jourMode}}{}{
 T-naive is distinct from ordinary T symmetry, in that it doesn't exchange
initial and final states, while still reversing the sign of momenta and
angular momenta. Despite this difference, studying the former with triple
T-odd product asymmetries, that are measurable in the considered baryonic
B decay, could provide insight into the latter~\cite{ref:Geng_Hsiao_JMPA}.
}

%
%
%
%
%

\ifthenelse{\boolean{confFlag}}{
\section{THE \babar\ DETECTOR AND DATA SET}
}{
\section{Dataset and Selection}
}

\label{sec:Dataset}

The data sample consists of $\TotalBPairs\times 10^6$ $B\bar{B}$ pairs,
corresponding to an integrated luminosity of \TotalOnPeakLumi\ ${\rm
fb^{-1}}$, collected at the \FourS\ resonance with the \babar\ detector.
The detector is described in detail elsewhere~\cite{ref:BaBar_NIM}.  
Charged-particle trajectories are measured in a tracking system consisting
of a five-layer double-sided silicon vertex tracker (SVT) and a 40-layer
central drift chamber (DCH), both operating in a 1.5-T axial magnetic
field.  A ring-imaging Cherenkov detector (DIRC)  is used for
charged-particle identification. A CsI(Tl) electromagnetic calorimeter
(EMC) is used to detect and identify photons and electrons, while muons
and hadrons are identified in the instrumented flux return of the magnet
(IFR). A \babar\ detector Monte Carlo simulation based on
\geantfour~\cite{ref:geantfour} is used to optimize selection criteria and
determine selection efficiencies.

\ifthenelse{\boolean{confFlag}}{
\section{EVENT SELECTION}
\label{sec:EventSelection}
}{
}

We reconstruct $\bar{\Lambda}$ candidates in the $\bar{\Lambda}\rightarrow
\bar{p}\pi$ decay mode as combinations of oppositely charged tracks,
assign the proton and pion mass hypotheses, and fit to a common vertex
\cite{ftn:Charge_Implied}.
Combinations with invariant mass in the range \LambdaInvMassCutL\ --
\LambdaInvMassCutU\gevcc\ are refit requiring the track pairs to originate
from a common vertex and constraining the mass to the world-average
$\Lambda$ mass \cite{ref:PDG}. Candidate $\B^0$ mesons are formed by
combining $\bar{\Lambda}$ candidates with two additional oppositely
charged tracks, each with momentum transverse to the beam greater than
\GoodTracksLoosePtMin\mevc.

Measurements of the average energy loss (\dedx) in the tracking devices,
the angle of the \Cherenkov\ cone in the DIRC, and energies deposited in the
EMC and IFR are combined to give a likelihood estimator $L_\alpha$
for a track to be
consistent with a given particle hypothesis $\alpha$.  We require that the
$\Lambda$-decay proton candidates satisfy the particle-identification
criteria $L_p/L_K>\PLHVeryLoosePKLikeRatioMin$ and
$L_p/L_\pi>\PLHVeryLoosePPiLikeRatioMin$ to discriminate from kaons and
pions, respectively. The candidate protons, which are assumed to originate
from the $B^0$ decay vertex, are analyzed with a selection algorithm based
on bagged decision trees~\cite{ref:Breiman_Bagging} which provide
efficient particle discrimination, retaining
\PBmcsigCumKMSuperLooseEffNoErr\ of the signal candidates and
\PBbkgKMSuperLooseEffNoErr\ of the background. The candidate pions from
the $B^0$ vertex are required to pass a similar selection algorithm, tuned
to discriminate pions, that retains \PiBmcsigCumKMSuperLooseEffNoErr\ of
the signal and \PiBbkgKMSuperLooseEffNoErr\ of the background. A Kalman
fit~\cite{ref:Hulsbergen_Kalman} to the full decay sequence is used to
reconstruct the $B^0$ vertex using the position of the beam spot and the
total beam energy as kinematic constraints. Only candidates with a fit
probability $P_{\rm vtx}>\VtxProbCutL$ are considered, a requirement that
retains \VtxCutmcsigEffNoErr\ of the signal and \VtxCutbkgEffNoErr\ of the
background.

The primary background arises from light-quark continuum events
$e^+e^-\rightarrow q\bar{q}$ ($q = u,d,s,c$), which are characterized by
collimation of final-state particles with respect to the quark direction,
in contrast to the more spherical $B\bar{B}$ events. Exploiting this shape
difference, we increase the signal significance using event-shape
variables computed from the center-of-mass (CM) momenta of 
charged and neutral particles in the event. For each event,
we combine the sphericity~\cite{ref:Hanson_Jets}, the angle
between the $B^0$ thrust axis and detector longitudinal axis, and the
zeroth and second-order Legendre polynomial moments
\cite{ftn:LegMoments}
of the tracks not associated
with the reconstructed B candidate,
into a Fisher discriminant~\cite{ref:Fisher}, where the coefficients are chosen to optimize the
separation between signal and continuum-background Monte Carlo samples. We
find that the selection using the optimal cut on the Fisher discriminant
retains
\FishermcsigEffNoErr\ of the candidates from the signal Monte Carlo sample
and \FisherbkgEffNoErr\ from the continuum-background Monte Carlo sample.

To further reduce the combinatoric background, we take advantage of the long
mean lifetime of $\Lambda$ particles and require that the separation of
the $\Lambda$ and $B^0$ vertices, divided by its measurement error,
computed on a per-candidate basis by the fit procedure, exceeds
\LambdaFlightSigCutL. This criterion is optimized on Monte Carlo events
and is effective in rejecting \LmFlCutdataSideRejNoErr\ of combinatoric
background that survives all other selection requirements, while retaining
\LmFlCutmcsigSideEffNoErr\ of the signal candidates. The only sizable
$B^0$ background is from the process $B^0\rightarrow
\bar{\Lambda_c}^-p\rightarrow\bar{\Lambda}p\pi^-$,
which we suppress by removing candidates with an invariant mass
$m(\bar{\Lambda}\pi^-)$ within \LambdaCInvMassCutSigma\ standard
deviations (\LambdaCInvMassCut\mevcc) of the nominal $\Lambda_c$ mass
\cite{ref:PDG}.

The kinematic constraints on $B^0$ mesons produced at the \FourS allow
further background discrimination from the variables \mes\ and \deltaE.
We define $ \mes =
\sqrt{\left(\frac{s}{2}+\vec{p}_i\cdot\vec{p}_B\right)^2/E_i^2 -
\left.\vec{p}_B\right.^{2}}$, where $(E_i,\vec{p}_i)$ is the four momentum
of the initial $e^+e^-$ system and $\vec{p}_B$ is the momentum of the
reconstructed $B^0$ candidate, both measured in the laboratory frame, and
$s$ is the square of the total energy in the $e^+e^-$ center-of-mass
frame. We define $ \deltaE = E^*_B - \frac{\sqrt s}{2} $, where $E^*_B$ is
the $B^0$ energy in the $e^+e^-$ center-of-mass frame. Signal candidates
have \mes\ close to the $B^0$ mass and \deltaE\ near zero.
Candidates satisfying $|\deltaE|<\DeltaEFitRegion\mev$ and
$\MesFitRegionLow<\mes<\MesFitRegionUp\gevcc$ are used in the
maximum-likelihood fitting process.

%
%
%
%
%

\ifthenelse{\boolean{confFlag}}{
\section{BRANCHING FRACTION}
}{
\section{Branching Fraction}
}

\label{ssec:sPlot_fit}

We measure the branching fraction with a maximum-likelihood fit on the \mes-\deltaE\ 
observables of reconstructed $B^0$ candidates. The \splot\ 
technique\ifthenelse{\boolean{physMode}}{ \cite{ref:Pivk_sPlot} }{
\cite{ref:Pivk_sPlot,ref:Pivk_sPlot_article,ref:Cahn_sPlot,ref:Snyder_sPlot}
} is then used to determine the \mlbp\ distribution and,
after correcting for the nonuniform
reconstruction efficiency, measure the \mlbp
-dependent differential branching fraction.

We consider as signal candidates only reconstructed $B^0$ candidates in
which all particles are correctly assigned in the decay chain. By
self-cross-feed, we refer to events in which $B^0$ mesons decay to
$\bar{\Lambda}p\pi$ and are reconstructed as signal candidates in which
one or more particles are not correctly assigned in the decay chain.
An example of such a misreconstruction is where the
protons from the signal $B^0$ and $\Lambda$ decays are interchanged.
We define the probability density function (\pdf) in the \deltaE-\mes\ plane as the sum
of signal, self-cross-feed, and background components. The likelihood 
function is given by
 \begin{align*}
{\like}=\frac{1}{N!}e^{-(N_{S} + N_{\rm scf} + N_{B})}
\prod_{e=1}^{N} 
\left\{N_{S}{\cal{P}}_{S}(y_e)
+ N_{\rm scf}{\cal{P}}_{\rm scf}(y_e)
\ifthenelse{\boolean{confFlag}}{}{
\right.\\\left.}
+ N_{B}{\cal{P}}_{B}(y_e)
\right\},\ \ \ (1)
\addtocounter{equation}{1}
\end{align*}
where $y_e=\left(\mes_{,e},\deltaE_e\right)$, the product is over 
the $N$ fitted candidates with $N_{S}$ and
$N_{B}$ representing the numbers of signal and background events,
and $N_{scf}\equiv N_S f_{\rm scf}$ representing the self-cross-feed
contribution. The three
$\cal{P}$ functions are taken as products of one-dimensional \deltaE\ and
\mes\ \pdf s. We are justified in this simplification by the small
correlation between these two variables in our Monte Carlo sample.
 The \mes\ \pdf\ is taken as a sum of two Gaussians for the signal and an
ARGUS function \cite{ref:Argus} for the background. The \deltaE\ \pdf\
is taken as a sum of two Gaussians for the signal and a first-order
polynomial for the background. Finally, the self-cross-feed contribution
shows a peaking component that is modeled as the product of a sum of two
Gaussians in \deltaE, and a single Gaussian in \mes.
We determine 
 $f_{\rm scf} = \mbox{\SCFExpectedFractionNoErr}$ and the other
parameters that characterize this background  from 
fits to simulated events.

We fit the means of the narrow \deltaE\ and \mes signal Gaussians, the
coefficient in the exponential of the Argus function, the linear
coefficient of the \deltaE\ background distribution, and the event yields
$N_S$ and $N_B$. The means of the wide Gaussians are
determined by applying Monte Carlo-determined offsets to the means of the
narrow ones, such that only an overall shift of the fixed \pdf\ shape is
allowed. All other parameters used in the likelihood definition
are fixed to values determined from fits to Monte Carlo-simulated events.

Once the maximum-likelihood fit provides the best estimates of the \pdf\
parameters, we use the \splot\ technique
to reconstruct the efficiency-corrected \mlbp\ distribution and measure
the branching fraction.  The \pdf\ is used to compute the 
s-weight for the $n$th component of event $e$ as
 \begin{equation}
	\mbox{}_{s}{\cal{P}}_n(y_e) =
\frac{\sum_{j=1}^{n_{c}}{{\bf V}_{nj} {\cal P}_{j}(y_e)}}
{\sum_{k=1}^{n_{c}}{{N}_{k} {\cal P}_{k}(y_e)}},
\label{eqn:splot_weights}
\end{equation}
 where the indices $n,j$, and $k$ run over the $n_{c}=3$ signal,
background, and self-cross-feed components. The symbol ${\bf V}_{nj}$ is the
covariance matrix of the event yields as measured from the fit to the data
sample. An important property of the \splot\ is that the sum of the s-weights
for the signal or background component equals the corresponding number of
fitted signal or background events. We have demonstrated with
simulated experiments that the \splot~is an unbiased
and nearly optimal estimator of the \mlbp\ distribution. To retrieve the
efficiency-corrected number of signal events in a given \mlbp\ bin $J$ we
use the s-weight sum
 \begin{equation}
	\tilde{N}_{S,J} = \sum_{e\in
J}{\frac{\mbox{}_{s}{\cal{P}}_S(y_e)}{\varepsilon(x_e)}}
 \label{eqn:eff_corrected_n},
\end{equation}
 where the per-event efficiency $\varepsilon(x_e)$ depends on the position
$x_e=\left(m_{\bar{\Lambda}p},\cos\theta_{\bar{\Lambda}\pi}\right)$ in
the square Dalitz plane. Here $\theta_{\bar{\Lambda}\pi}$ is the angle
between the momenta of the pion and the $\bar{\Lambda}$ candidate in the
$\bar{\Lambda} p$ rest frame, and the efficiency is determined over a
\DalitzBoxN\ grid in the Dalitz plane, using fully reconstructed
signal-Monte Carlo events.  The error $\sigma [\tilde{N}_{S,J}]$ in
$\tilde{N}_{S,J}$ is given by
 \begin{equation}
\sigma^2 [\tilde{N}_{S,J}] =
\sum_{e\in J}{\left(\frac{\mbox{}_{s}{\cal{P}}_S(y_e)}{\varepsilon(x_e)}
\right)^2}.
\end{equation}
 An estimate of the efficiency-corrected number of signal events in the
sample is given by the sum of the efficiency-corrected s-weights, or
 \begin{equation}
	\tilde{N}_S = \sum_{J}{\tilde{N}_{S,J}}\ ,
\end{equation}
 and the branching fraction is obtained from 
\begin{equation}
{\cal{B}}\left(\blppi\right)
	= \frac{\tilde{N}_S}{N_{B\bar{B}}\cdot
{\cal{B}}\left(\Lambda\rightarrow p\pi\right)},
\label{eqn:bratio_expr}
\end{equation}
where $N_{B\bar{B}}$ is the total number of $B\bar{B}$ pairs
and ${\cal{B}}\left(\Lambda\rightarrow p\pi\right)=\LambdaPPiBF$ \cite{ref:PDG}.
 Using a collection of Monte Carlo pseudoexperiments, in which signal
candidates, generated and reconstructed with a complete detector
simulation, were mixed with background candidates, generated according to
the background \pdf, we confirm that this procedure provides a measurement
of the branching fraction with negligible biases
and accurate errors.

We can measure the CP-violating branching-fraction asymmetry by tagging
the flavor of the $B^0$ ($\bar{B^0}$) meson with the charge of its daughter proton
(antiproton). We repeat the maximum-likelihood fit described above including
the partial rate asymmetry
 \begin{equation}
{\cal{A}} = \frac{{\cal{B}}({\bbarlppi})-{\cal{B}}({\blppi})}
{{\cal{B}}({\bbarlppi})+{\cal{B}}({\blppi})}
\end{equation}
 as a free parameter. We reduce the effect of systematic differences in
particle-identification efficiencies between protons and antiprotons, and
between positive and negative pions, by performing the fit on a sample of
reconstructed candidates, where protons and pions that originate from the
$\bar{\Lambda}$ decay satisfy the same particle-identification criteria as
those imposed on the protons and pions that originate from the $B^0$ vertex.

%
%
%
%
%

\ifthenelse{\boolean{physMode}}{}{
\clearpage
}

\section{$\bar{\Lambda}$ Polarization Measurement}

\label{sec:Helicity_measurement}

We study the three orthogonal components of the polarization of
$\bar{\Lambda}$ candidates reconstructed in the \blppi\ decay as a
function of \ELambdabar, the $\bar{\Lambda}$ energy in the $B^0$ rest
frame \cite{ref:Suzuki_JPG29}. The distribution of the helicity angle
\ThetaH\ for the $\bar{\Lambda}$ decay is given by
 \begin{equation}
\frac{1}{\Gamma} \frac{d\Gamma}{d \CosThetaH} 
= \frac{1}{2} \left[ 1 + 
\alpha_{\bar{\Lambda}}P\left(\ELambdabar\right) \CosThetaH\right],
\label{eq:HelicitySignalDistrib}
\end{equation}
 where \ThetaH\ is the angle between the antiproton direction, in the
$\bar{\Lambda}$ rest frame, and either (1) $\hat{L}$, the unit vector in
the direction of the $\bar{\Lambda}$ in the $B^0$ rest frame; (2)
$\hat{T}$, the unit vector  along the
direction of the cross product between the momenta,  in the $B^0$ rest frame,
of the proton and the $\bar{\Lambda}$;
or (3) $\hat{N}= \hat{L} \times \hat{T}$.
The symbol $P\left(\ELambdabar\right)$ is the component of the $\bar{\Lambda}$
polarization in the $\hat{L}$, $\hat{T}$, or $\hat{N}$ direction
as a function of \ELambdabar, and $\alpha_{\bar{\Lambda}}$ is the
$\bar{\Lambda}$ decay-asymmetry parameter \cite{ref:PDG}. \CP\
conservation in \blppi\ decays implies that
 \begin{equation}
\alpha_{\Lambda}P_{[L,N],\bar{B}^0\rightarrow\Lambda\bar{p}\pi} 
\left(\ELambdaB\right) 
=\alpha_{\bar{\Lambda}}P_{[L,N],B^0\rightarrow\bar{\Lambda}p\pi}
\left(\ELambdabar\right),
\end{equation}
 while the product $\alpha_{\Lambda}P_T$ changes sign under \CP\
conjugation. We use these relations to fit the $B^0$ and $\bar{B^0}$
candidate samples together.

\begin{figure}[t]
\mbox{
\includegraphics[width=8.0cm]
{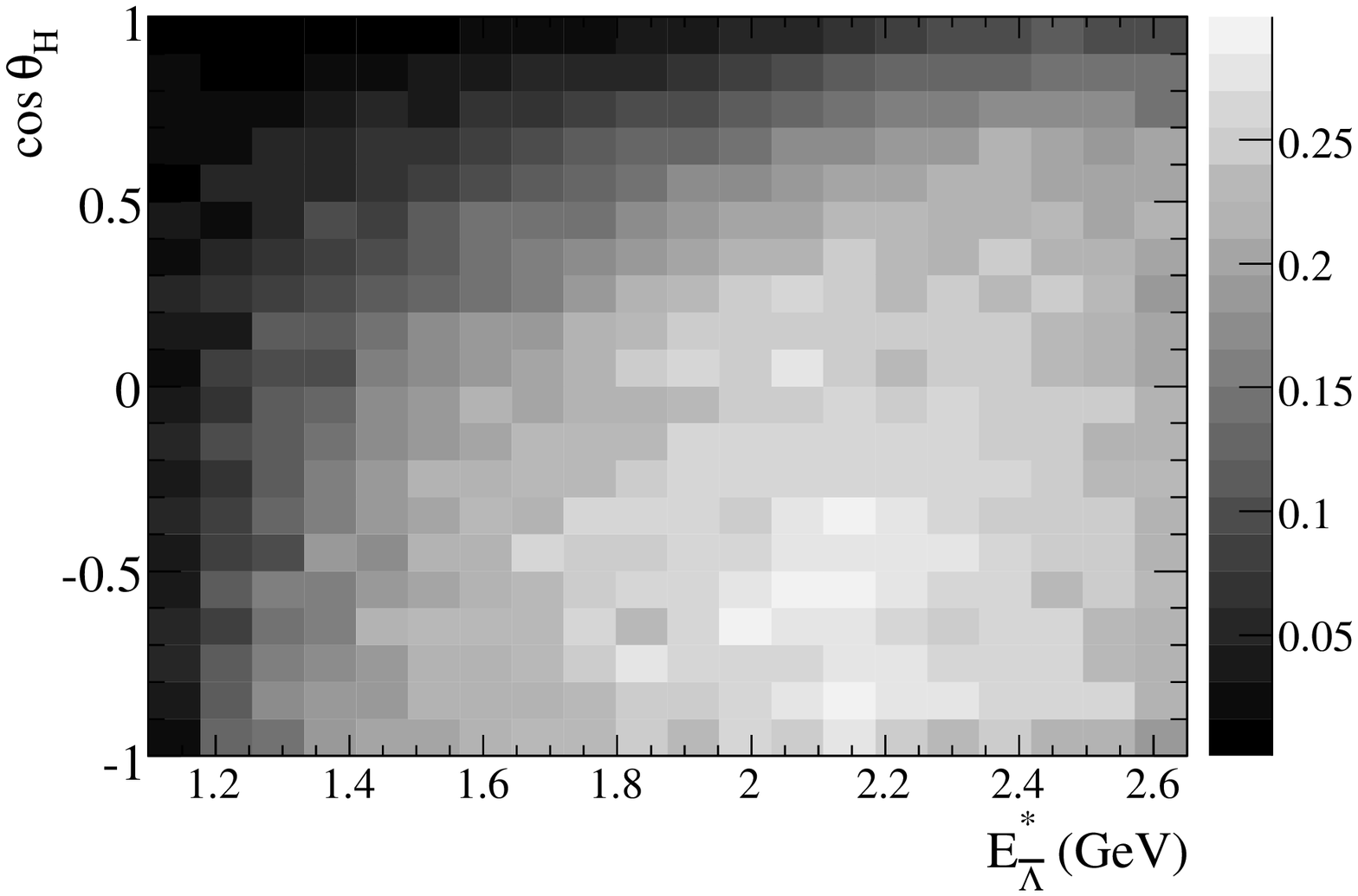}}
 \caption{Reconstruction efficiency measured on the Monte Carlo signal
sample, as a function of \CosThetaH\ and \ELambdabar. In this plot \ThetaH\ 
is the angle between the antiproton direction, in the
$\bar{\Lambda}$ rest frame, and $\hat{L}$, the unit vector in
the direction of the $\bar{\Lambda}$ in the $B^0$ rest frame.}
 \label{fig:ELambdaBCosThetaH_Efficiency}
\end{figure}

We use a maximum-likelihood fit in \mes, \deltaE, \ELambdabar, and
\CosThetaH\ to measure the polarization as a function of \ELambdabar\
along each of the three axes defined above. We divide the \ELambdabar\
range into \ElbMaxLikeBinNumberLett\ bins with boundaries
\ElbMaxLikeBinZeroLow, \ElbMaxLikeBinOneLow, \ElbMaxLikeBinTwoLow, and
\ElbMaxLikeBinTwoHigh\ \gev, chosen in order to have similar numbers of
signal events in each bin. We define a \pdf\ as the sum of signal and
background components. The likelihood is
 \begin{eqnarray*}
{\like}&=&\frac{1}{N!}
\prod_{k=1}^{3}e^{-\left(N_{k,S} + N_{k,B}\right)}\prod_{e=1}^{N_k}
 \left[N_{k,S}{\cal{P}'}_{k,S}(z_e){\cal{P}}_S(y_e)+\right.\\
&&\hspace{8.7em}+\left.N_{k,B}{\cal{P}'}_{k,B}(z_e){\cal{P}}_B(y_e)\right],
\ (10)
\addtocounter{equation}{1}
 \end{eqnarray*} 
 where we have divided the observables into two sets $y_e=(\mes,\deltaE)$
and $z_e=(\CosThetaH,\ELambdabar)$, and the products are over the three bins
in \ELambdabar\ and over the $N_k$ events that populate the $k$th
bin, where $N_{k,S}$ and $N_{k,B}$ represent the numbers of fitted signal
and background events. The ${\cal{P}}_{S,B}(y_e)$ \pdf s are the same
functions used in the branching-fraction measurement. However the
self-cross-feed component is not included since it corresponds to a
negligible fraction of the signal events. For the $k$th bin in
$\ELambdabar$, the signal (\CosThetaH,\ELambdabar)  \pdf\ is written as
the product of the differential branching fraction of
Eq.~\ref{eq:HelicitySignalDistrib}, times the signal-reconstruction
efficiency $\epsilon\left(\ThetaH,\ELambdabar\right)$:
 \begin{equation} 
{\cal{P}'}_{k,S}\left(\ThetaH,\ELambdabar\right)=
\frac{1}{2}\epsilon\left(\ThetaH,\ELambdabar\right) \left[ 1 +
\{\alpha_{\bar{\Lambda}}P\}_k \CosThetaH\right]
 \end{equation} 
 where the $\{\alpha_{\bar{\Lambda}}P\}_k$ are fit parameters. The
signal-selection efficiency is measured with a sample of reconstructed
signal-Monte Carlo events that pass the same selection criteria as those used to
define the data sample. We bin the signal efficiency in
\ElbCosLHeliEffBoxN\ rectangular boxes that cover the allowed region of
the \ELambdabar-\CosThetaH\ plane (Fig.~\ref{fig:ELambdaBCosThetaH_Efficiency}).

The background \ThetaH\ distribution is modeled as a linear combination of
\Chebyshev\ polynomials up to \HeliBkgChebyDegree\ order. The
\HeliBkgChebyCoeffN\ coefficients that define the linear combination are
fitted independently for each of the \ElbMaxLikeBinNumberLett\ bins in
\ELambdabar. We study the \ThetaH\ distribution of background events using
candidates in the sideband region $\mes<\HeliMesSideband\gevcc$,
 and find it to be nearly independent of \mes. We
consider this insensitivity as an indication that the shape of the
background \ThetaH\ distribution is the same for events in and out of the
signal region.

We have confirmed that this \pdf\ representation does not bias the
polarization measurement by performing pseudoexperiments in which signal
candidates, generated and reconstructed with a complete detector
simulation, were mixed with background candidates generated according to
the observed helicity distribution in the $\mes<\HeliMesSideband\gevcc$
sideband. The number of signal and background candidates are chosen to match the
characteristics of the data.

%
%
%
%
%

\ifthenelse{\boolean{confFlag}}{
\section{SYSTEMATIC UNCERTAINTIES}
}{
\section{Systematic uncertainties}
}
\label{sec:Systematic_errors}

Systematic uncertainties in the branching-fraction measurement are listed
in Table I and classified as overall uncertainties, uncertainties
associated with event selection, and uncertainties associated with fitting
the event distribution. We study the uncertainty due to tracking
efficiency by comparing data and Monte Carlo for a sample of $\tau$-pair
events, in which one $\tau$ decays to one charged track and the other
$\tau$ decays to three charged tracks. We separately study the tracking
efficiency of $\Lambda$ decay products using an inclusive sample
of $\Lambda\rightarrow p \bar{\pi}$ candidates
and estimate an overall tracking reconstruction efficiency of
\TotalTrackSystematic. The uncertainty associated with
particle-identification (PID) selection criteria is estimated as
\PIDSystematic\ by comparing data and Monte Carlo identification
efficiencies for protons from $\Lambda\rightarrow p \pi$ decays and pions
from $K^0_S\rightarrow \pi\pi$ decays. The limited signal-Monte Carlo
sample available to measure the reconstruction efficiency over the Dalitz
plane results in an additional \DalitzStatSystematic\ contribution. The
uncertainty in the number of $B\bar{B}$ pairs in the data sample accounts
for a \TotalBPairsPerc\ contribution, while the assumption of a
\BZeroOverBChargedDefault\ ratio of $B^0\bar{B^0}$ to $B\bar{B}$ at the
\FourS\ gives an additional \BFractionSystematic\ contribution, computed
from the difference between 50\% and the current measured value
\BZeroOverBChargedFraction\ \cite{ref:PDG}.

Uncertainties associated with event-selection requirements on the Fisher
discriminant and vertex fit probability are estimated by
comparing data and Monte Carlo-selection efficiencies for a sample of
$B^0\rightarrow J/\psi K^0_S$ candidates. We use an inclusive sample of
$\Lambda\rightarrow p\pi$ candidates to estimate uncertainties associated
with the efficiencies of the flight-length significance and
$\Lambda$-mass requirements.

The application of the requirement on the reconstructed $m(\Lambda\pi)$
invariant mass to suppress $B^0\rightarrow\bar{\Lambda_c} p$ background
has two associated systematic effects. The first results in an approximate
\LambdaCPiResBackSystematic\ increase in the branching fraction due to the
residual $\Lambda_c$ component that survives the cut. The second results
in an approximate \BLambdaPPiVetoSystematic\ reduction of the branching
fraction due to the reduced Dalitz-plot phase space.  We correct for these
effects and take the larger of the two as the uncertainty associated with
the $\Lambda_c$ veto cut.

\begin{table}[t]
\caption[Breakdown of systematic errors]{Systematic
uncertainties on the branching-fraction measurement. ``Total'' is the
sum in quadrature of all the individual contributions.}

\begin{center}
\begin{tabular}{c c c}
\hline
\hline
& Source & Uncertainty (\%)\\
\hline
Overall &Tracking efficiency & \TotalTrackSystematicNoPerc \\
&PID efficiency & \PIDSystematicNoPerc \\
&MC statistics & \DalitzStatSystematicNoPerc \\
&  $B\bar{B}$ counting & \TotalBPairsSystematicNoPerc \\
&$B^0\bar{B^0}/B\bar{B}$ fraction  & \BFractionSystematicNoPerc \\
&$\Lambda\rightarrow p\pi$ branching fraction & 
\LambdaPPiBFSystematicNoPerc \\
\hline
Event selection&Event shape  & \FDCutSystematicNoPerc\\
requirements &Fit probability &  \VtxCutSystematicNoPerc\\
&$\Lambda$ flight length & \LambdaFlightSystematicNoPerc\\
&$\Lambda$ mass & \LambdaMassSystematicNoPerc\\
& $\Lambda_c$ veto & \LambdaCVetoSystematicNoPerc\\
\hline
Fit procedure&Likelihood parameters & \LikelihoodParametersNoPerc\\
&\deltaE\ resolution& \DeltaESystematicNoPerc\\
&Self cross-feed fraction & \SCFSystematicNoPerc\\
&\splot\ bias & \LikelihoodBiasNoPerc\\

\hline
Total && \TotalSystematicNoPerc\\
\hline
\hline
\end{tabular}
\label{tab:systematics}
\end{center}
\end{table}

We vary parameters that are kept fixed in the likelihood fit by their
uncertainties, as measured on the signal-Monte Carlo sample fit, and measure the
variation of the \splot\ fitted result. The uncertainties associated with
the parameters that enter the definition of the signal \pdf\ are
conservatively considered as correlated and are thus added to give a
signal-\pdf\ overall uncertainty of \SignalPDFSystematic, where the
uncertainty in signal-\mes\ fixed parameters accounts for a
\MesSystematic\ contribution and that in signal-\deltaE\ fixed parameters
for a \DESystematic\ contribution. The same procedure is applied to the
parameters that enter the background \pdf, with uncertainties
determined on luminosity-weighted background-Monte Carlo samples, giving an
additional \BackgroundPDFSystematic\ uncertainty. Finally, we combine the
two uncertainties in quadrature and obtain a \LikelihoodParameters\
uncertainty associated with the shapes of the signal and background
models.  The comparison of $B\rightarrow J/\psi K_S^0$ data and Monte Carlo samples
reveals that the width of the \deltaE\ Gaussian in the signal \pdf\ can be
underestimated in the Monte Carlo by up to \DeltaEScaleDiffPerc, which
translates to an additional \DeltaESystematic\ uncertainty.

We estimate possible biases associated with the determination of yields
with the \splot\ technique, using an ensemble of Monte Carlo experiments.
Signal events, generated and reconstructed with a complete detector
simulation, were mixed with background events, generated according to
the background \pdf. The numbers of events were chosen according to the 
expected yields in the data sample under study. We estimate an
uncertainty of \LikelihoodBias.

The main systematic uncertainty in the polarization measurement is
associated with the limited statistics of the Monte Carlo sample used to
measure the signal-reconstruction efficiency in the
$(\CosThetaH,\ELambdabar)$ plane, which results in
$\alpha_{\bar{\Lambda}}P_L\left(\ELambdabar\right)$ uncertainties of
\MaxModulationPolarizSystematicA, \MaxModulationPolarizSystematicB, and
\MaxModulationPolarizSystematicC\ for the three \ELambdabar\ bins.
Variation of parameters fixed in the likelihood fit within their uncertainties
provides additional contributions
of \TotLikePolarizSystematicA, \TotLikePolarizSystematicB, and
\TotLikePolarizSystematicC\ in the three bins, respectively. We correct
the fit result for the small biases we observe in a sample of Monte Carlo
experiments, where background candidates were generated with the helicity
distribution observed in $\mes<\HeliMesSideband\gevcc$ sideband data, and
conservatively take these shifts as contributions to the systematic
uncertainty.

%
%
%
%
%

\ifthenelse{\boolean{confFlag}}{
\section{BRANCHING-FRACTION RESULTS}
}{
\section{Branching-Fraction Results}
}

\label{sec:BR_Results}

%
%

\ifthenelse{\boolean{confFlag}}{
\begin{figure}[t]
\begin{center}
\includegraphics[width=7.0cm]
{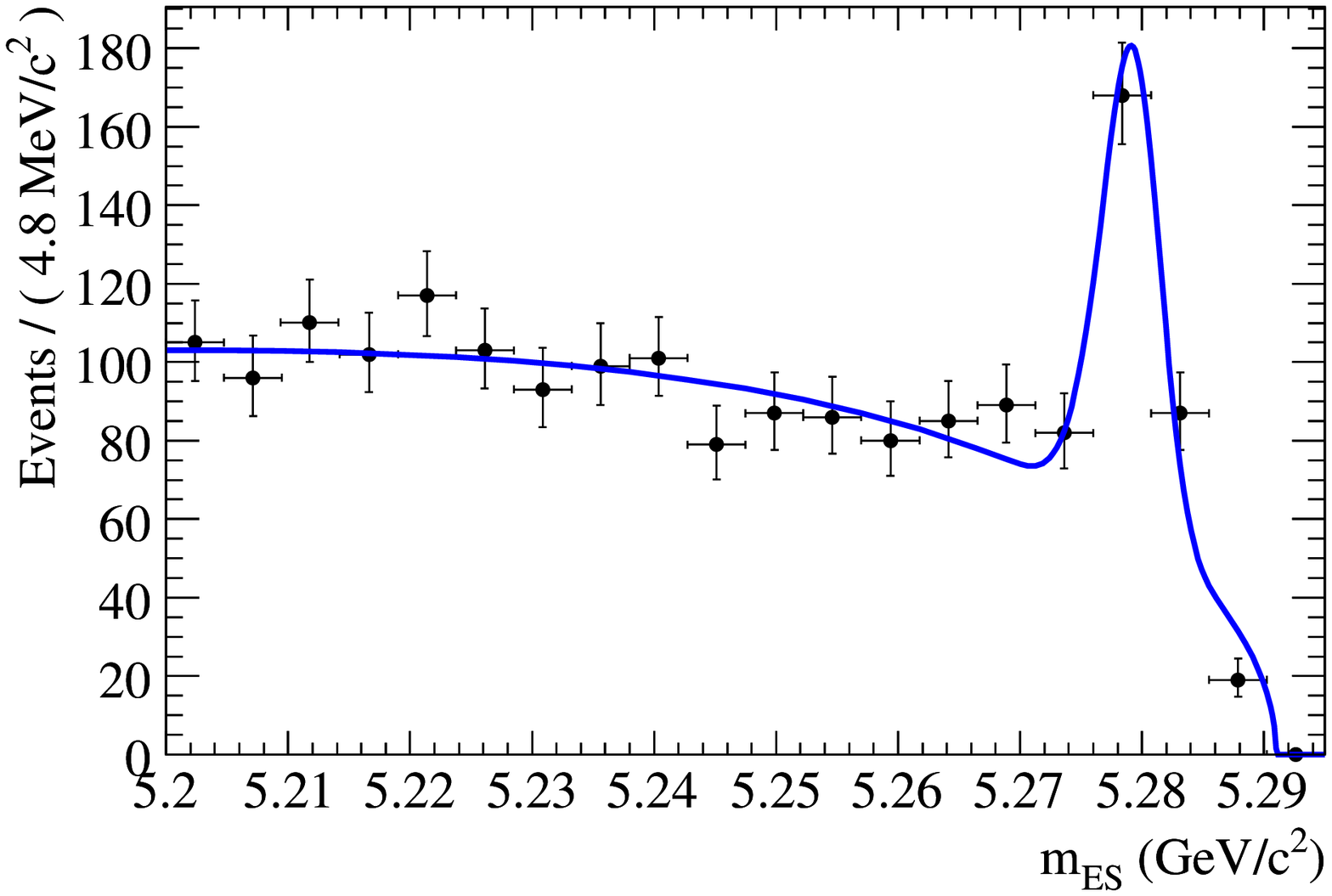}
\includegraphics[width=7.0cm]
{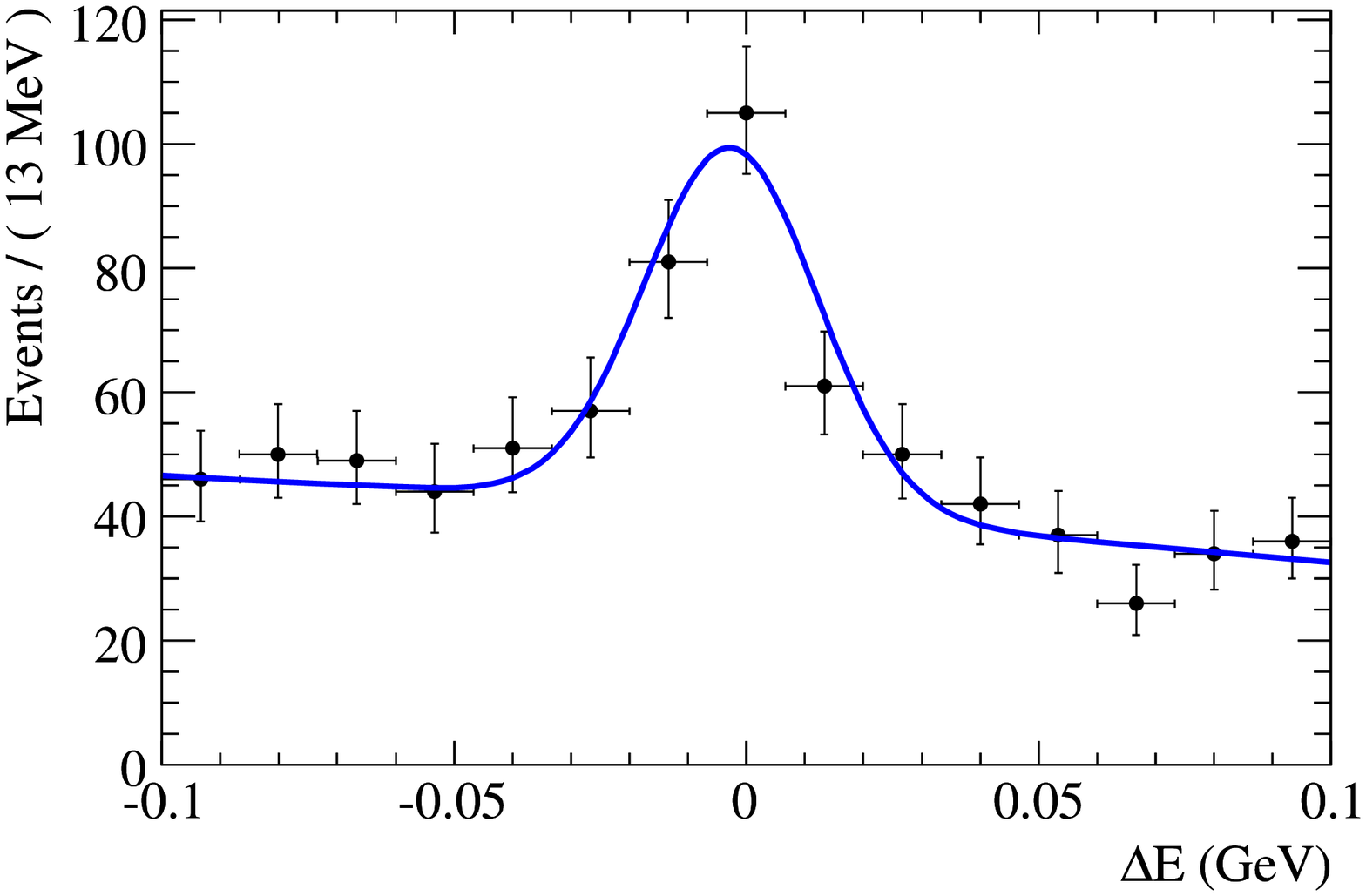}
\end{center}
\caption[Fit projections on \mes\ and \deltaE\ axes]{Left plot: 
\mes\ distribution of candidates with 
$|\deltaE|<\DeltaEWinCut\mev$. Right plot:
\deltaE\ distribution of candidates with
$\mes>\MesLowWinCut\gevcc$. The projections
of the two-dimensional fit \pdf\ are shown superimposed.}
\label{fig:Data_fit_Mes_DE_Projections}
\end{figure}
}{
\begin{figure}[t]
\begin{center}
\includegraphics[width=8.5cm]
{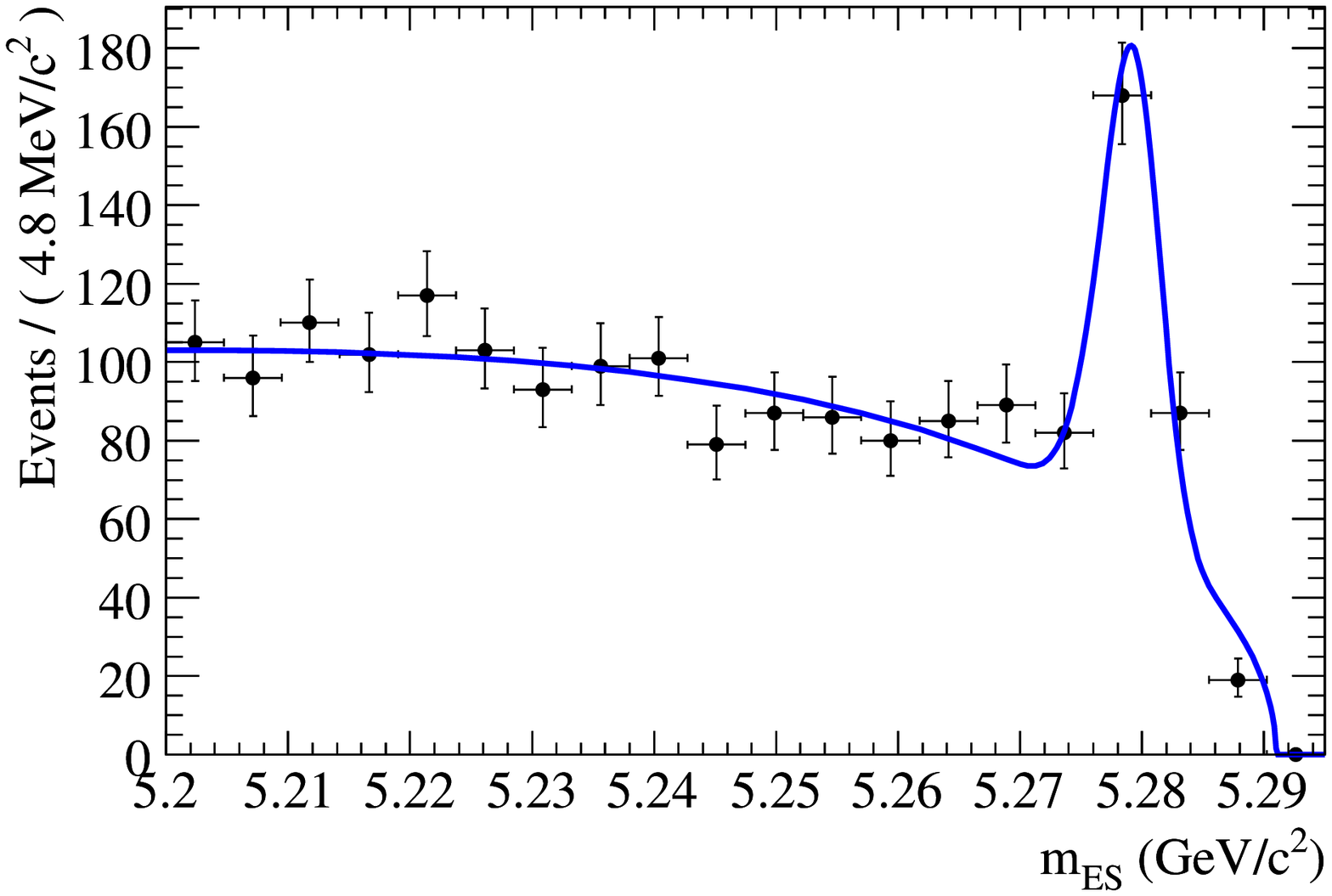}
\includegraphics[width=8.5cm]
{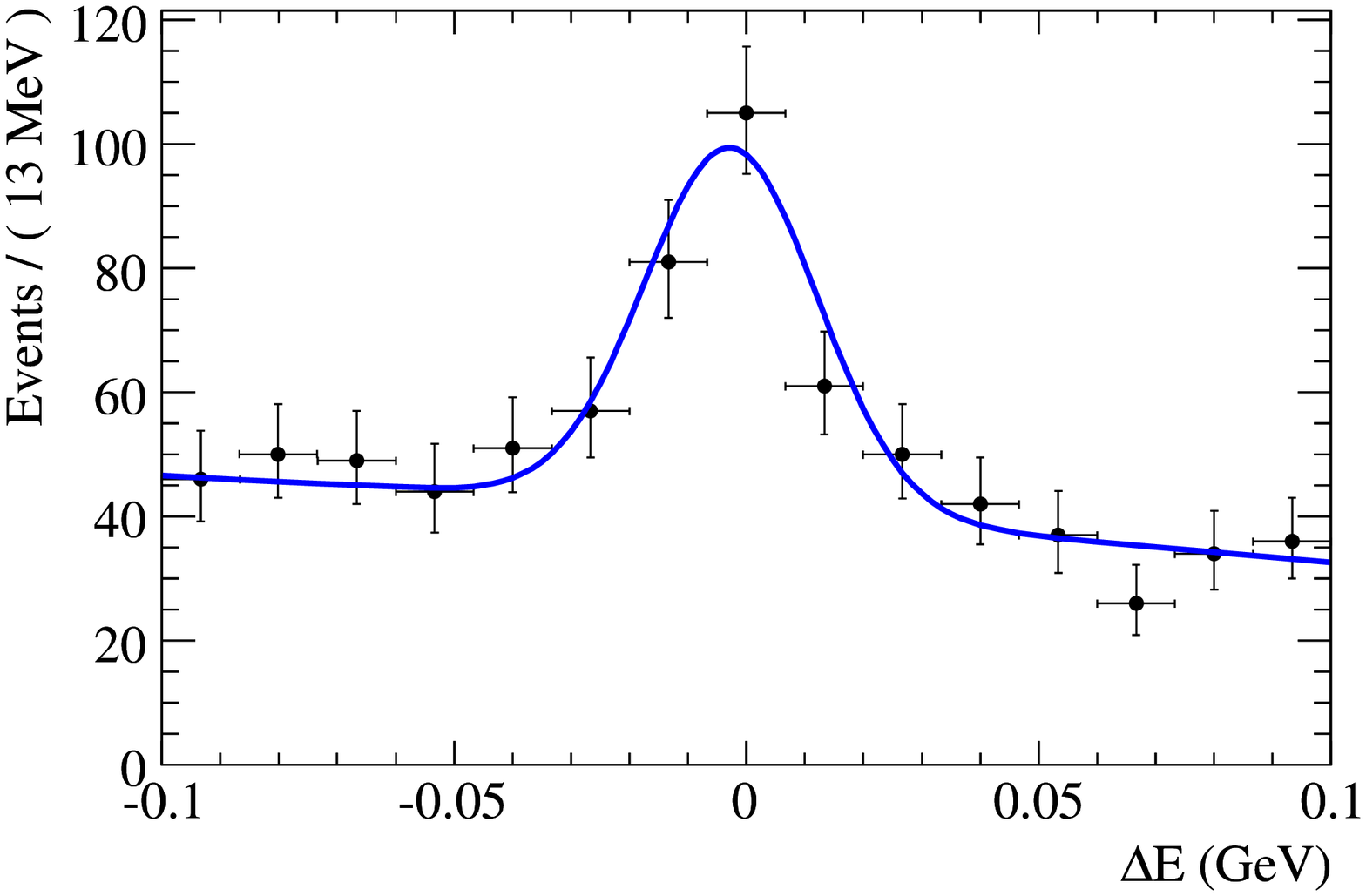}
\end{center}
\caption[Fit projections on \mes\ and \deltaE\ axes]{Upper plot: 
\mes\ distribution of candidates with 
$|\deltaE|<\DeltaEWinCut\mev$. Lower plot:
\deltaE\ distribution of candidates with
$\mes>\MesLowWinCut\gevcc$.  The projections
of the \mbox{two-dimensional} fit \pdf\ are shown superimposed.}
\label{fig:Data_fit_Mes_DE_Projections}
\end{figure}
}

\begin{figure}[t]
\begin{center}
\includegraphics[width=8.5cm]
{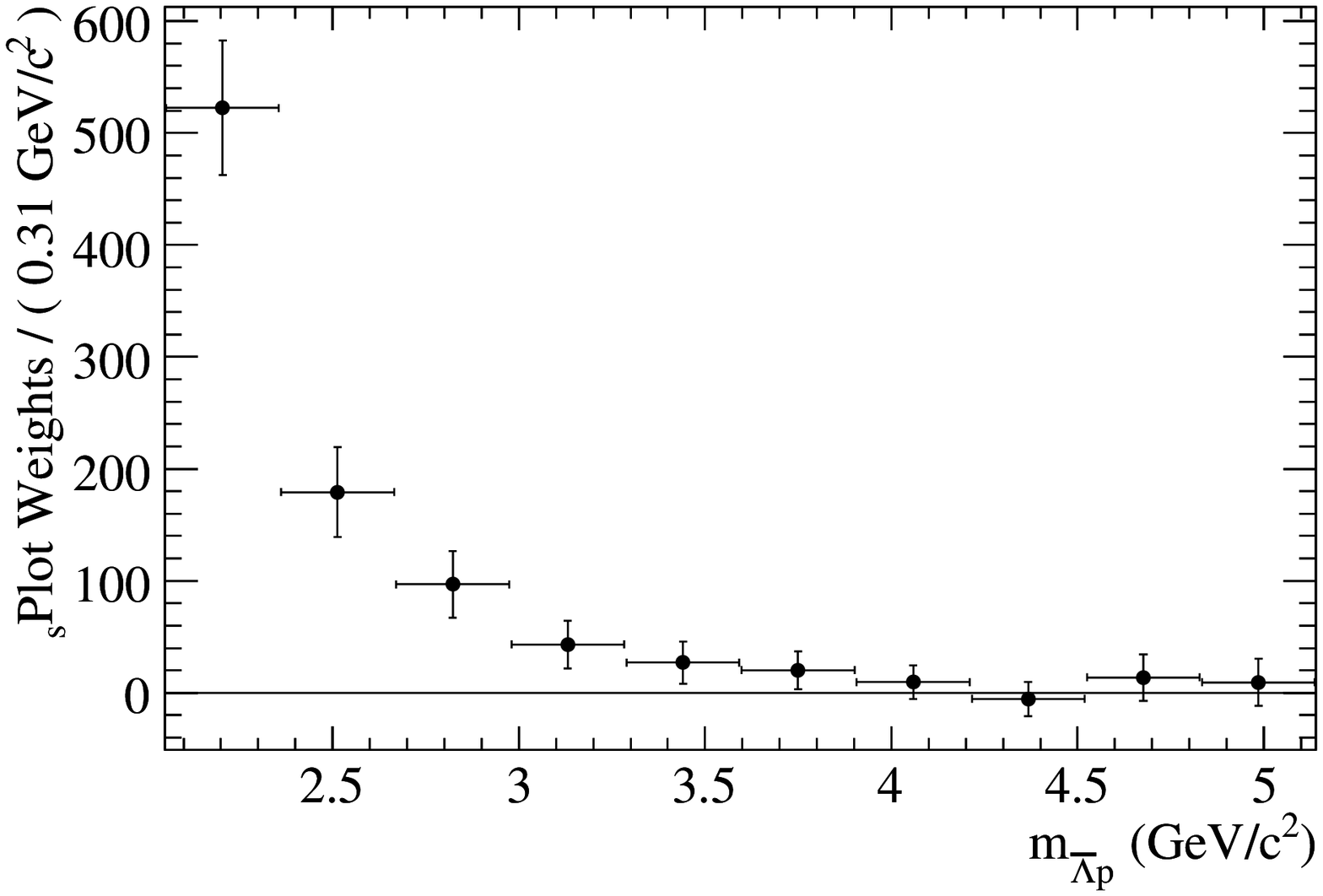}
\includegraphics[width=8.5cm]
{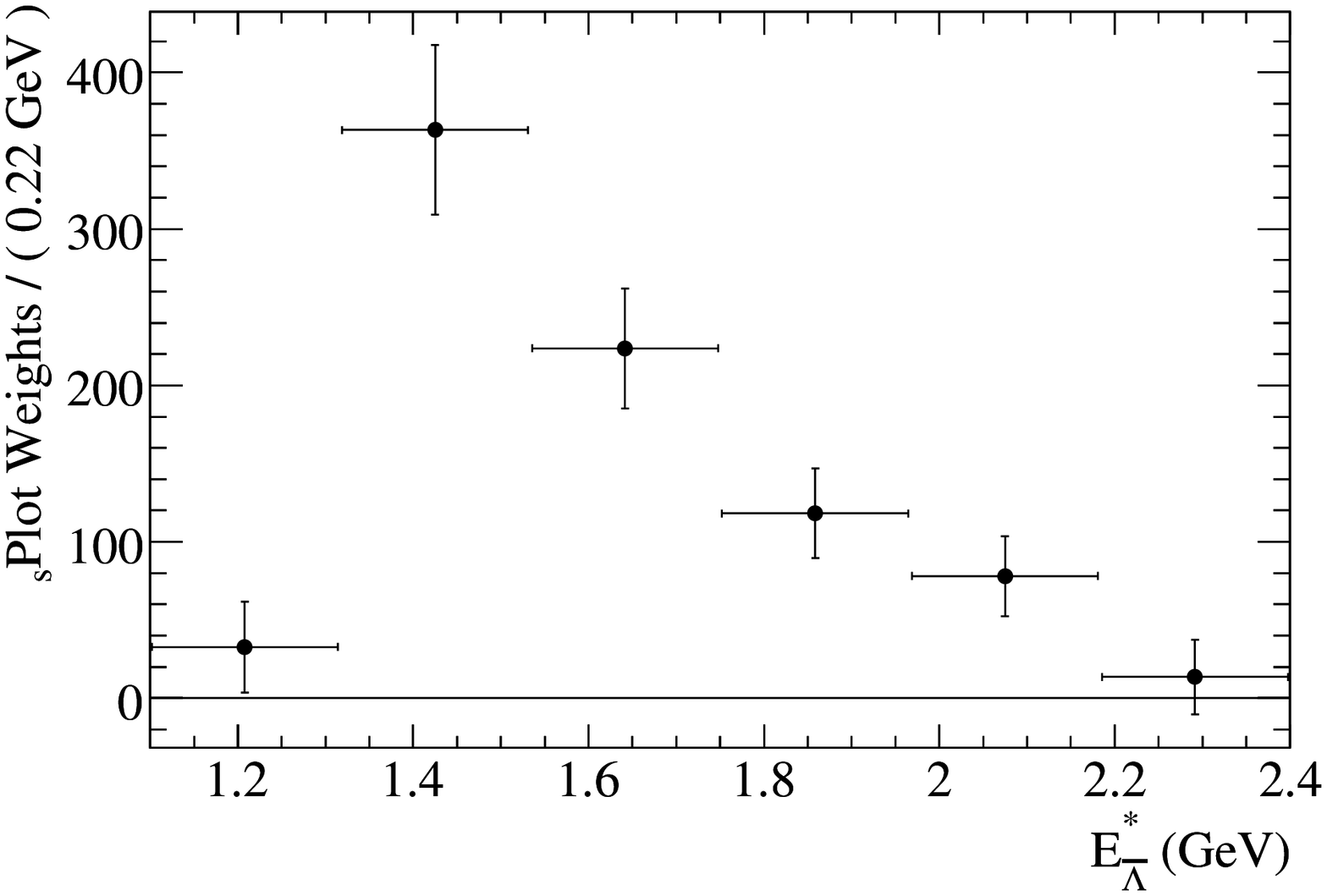}
\end{center}

\caption[\mlp\ event distributions]{Upper plot: \splot\ of the \mlp\ event
distribution with efficiency corrections
applied. Lower plot: \splot\ of the \ELambdabar\ distribution
with efficiency corrections applied.  Horizontal
bars represent bin ranges.}
\label{fig:Data_Mlp_Distributions}
\end{figure}

\begin{table}[t]
 \caption[Likelihood fit results]{Branching-fraction
results. $N_S$ and $N_B$ are the numbers of fitted signal and background
events, respectively. The symbol $\mu\left(\deltaE\right)$ is the mean for the
narrow Gaussian of the $\deltaE$ signal-\pdf\ component, while
$c_1\left(\deltaE\right)$ is the slope of the linear $\deltaE$ background
\pdf. $\mu\left(\mes\right)$ is the mean for the Gaussian of the
\mes\ signal \pdf, and $c_{\rm ARGUS}\left(\mes\right)$ is the coefficient
of the exponent in the background \mes\ Argus function \cite{ref:Argus}.
The uncertainties are statistical. }
 \begin{center}
\renewcommand{\tabcolsep}{2ex}
\renewcommand{\arraystretch}{1.5}
\begin{tabular}{c c}
\hline
\hline
Parameter & Value\\
\hline
$N_{S}$ & \LikeNSFitResult\\
$N_{B}$ & \LikeNBFitResult\\
\hline
$\mu\left(\deltaE\right)$ & \LikeMuDEFitResult \mev\\
$c_1\left(\deltaE\right)$ & \LikeCOneDEFitResult $\gev^{-1}$\\
\hline
$\mu\left(\mes\right)$ & \LikeMuMesFitResult\gevcc\\
$c_{\rm ARGUS}\left(\mes\right)$ & \LikeCArgusMesFitResult\\
\hline
\hline
\end{tabular}
\label{tab:LikelihoodFitResults}
\end{center}
\end{table}

We select a total of \DataSampleSelectedEvents\ candidates in the region
$|\deltaE|<\DeltaEFitRegion\mev$, $\mes>\MesFitRegion\gevcc$,
$|m(\Lambda\pi)-m(\Lambda_c)|>\LambdaCInvMassCut\mevcc$. Table
\ref{tab:LikelihoodFitResults} reports the fitted values of the
two-dimensional \mes-\deltaE\ \pdf\ parameters, while
Fig.~\ref{fig:Data_fit_Mes_DE_Projections} shows projections of the
two-dimensional \pdf\ on the \mes\ and \deltaE\ axes.
Figure~\ref{fig:Data_Mlp_Distributions} shows the efficiency-corrected
signal-\splot\ distribution of candidates as a function of \mlbp,
demonstrating a near-threshold enhancement similar to that observed in
other baryonic $B$ decays. Summing the content of the 
efficiency-corrected \splot\ bins,
we obtain $\mbox{\SPlotEffCorrectedYield}$ signal events, where the
uncertainty is statistical. Using Eq.~\ref{eqn:bratio_expr}, we measure the
branching fraction: $${\cal{B}}(B^0\rightarrow \bar{\Lambda} p \pi^-) =
\FinalBFCombined.$$ This measurement, which is compatible with a previous
measurement by the \belle\ collaboration \cite{ref:Belle_Blm_Update},
confirms the peaking of the baryon-antibaryon mass spectrum near
threshold, a feature that plays a key role in the explanation of the
larger branching fractions of three-body baryonic B decays compared to
two-body decays \cite{ref:Chua_Hou_EurC29}. From the maximum-likelihood
fit to the branching-fraction asymmetry we obtain: $${\cal{A}} =
\FinalAcpCombined,$$ which is compatible with zero asymmetry.

%
%

\begin{figure}[t]
\begin{center}
\includegraphics[width=8.5cm]
{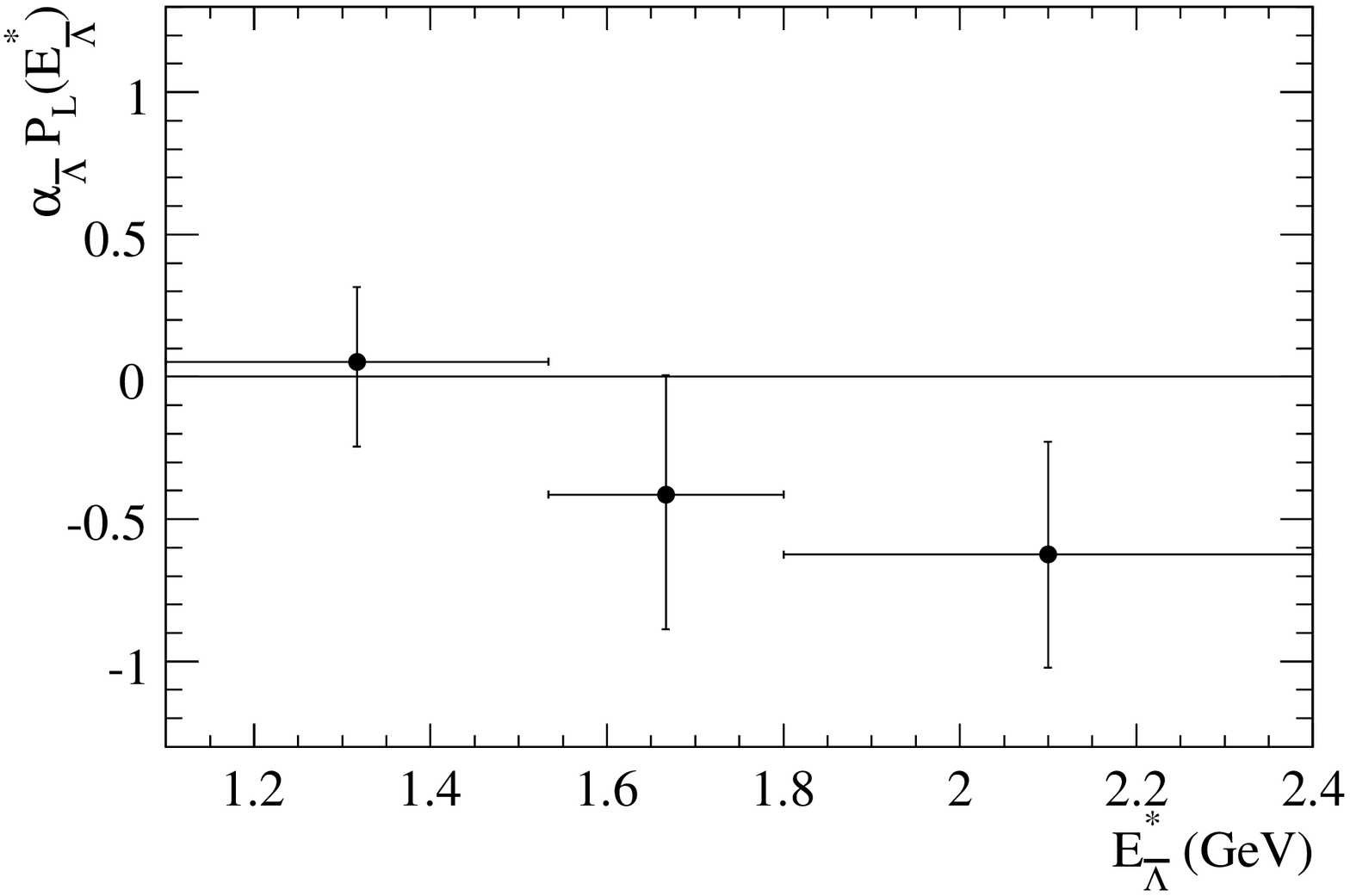}
\end{center}

\caption[\CosThetaH\ fit result]{The product of $\bar{\Lambda}$
longitudinal polarization and $\alpha_{\bar{\Lambda}}$ as a function of
\ELambdabar. Horizontal bars represent bin ranges.}
 \label{fig:Data_Helicity_Distribution}
\end{figure}

\ifthenelse{\boolean{confFlag}}{
\section{POLARIZATION RESULTS}
}{
\section{Polarization Results}
}

\label{sec:Pol_Results}

Only \DataSampleElbCutSelectedEvents\ candidates populate the \ELambdabar\
range $[\ElbMaxLikeRangeLow,\ElbMaxLikeRangeHigh]\gev$. Signal
candidates are absent in the region with $\ELambdabar >
\ElbMaxLikeRangeHigh$\ \gev\ (Fig.~\ref{fig:Data_Mlp_Distributions}) as a kinematical consequence of the near-threshold
peaking of the baryon-antibaryon mass spectrum.

We plot in Fig.~\ref{fig:Data_Helicity_Distribution} the values of the
longitudinal polarization product
$\alpha_{\bar{\Lambda}}P_L\left(\ELambdabar\right)$ obtained from the
maximum-likelihood fit. Table \ref{tab:HelicityFitResults} displays the
longitudinal, transverse, and normal polarization measurements in each of
the \ElbMaxLikeBinNumberLett\ \ELambdabar\ bins, assuming
$\alpha_{\bar{\Lambda}}=-\LambdaAlpha$ for the $\bar{\Lambda}$
decay-asymmetry parameter \cite{ref:PDG}. The results are
consistent with full longitudinal right-polarization of $\bar{\Lambda}$'s 
from \blppi\ decays at large \ELambdabar\
($\Lambda$'s would be oppositely polarized). The transverse polarization
is not expected to be zero because of the presence of
strong final-state interactions.

\begin{table}[t]
 \caption[Helicity fit results]{Polarization results. $N_S$ and $N_B$ are
the numbers of fitted signal and background candidates in each
\ELambdabar\ bin. We report the values of the longitudinal, transverse,
and normal $\bar{\Lambda}$ polarizations in each of the three \ELambdabar\
bins. }
 \begin{center}
\renewcommand{\tabcolsep}{0.65ex}
\renewcommand{\arraystretch}{1.5}
\begin{tabular}{c c c c}
\hline
\hline
 & \multicolumn{3}{c}{\ELambdabar\ range (\gev ) } \\

 & 
$\ElbMaxLikeBinZeroLow - \ElbMaxLikeBinOneLow$
& $\ElbMaxLikeBinOneLow - \ElbMaxLikeBinTwoLow$ 
& $\ElbMaxLikeBinTwoLow - \ElbMaxLikeBinTwoHigh$\\
\hline
$N_{S}$ & \HeliBinZeroNS & \HeliBinOneNS & \HeliBinTwoNS \\
$N_{B}$ & \HeliBinZeroNB & \HeliBinOneNB & \HeliBinTwoNB \\
\hline
$P_L$ & \HeliBinZeroPolariz &
\HeliBinOnePolariz & \HeliBinTwoPolariz \\
\hline
$P_T$ & \HeliTBinZeroPolariz &
\HeliTBinOnePolariz & \HeliTBinTwoPolariz \\
\hline
$P_N$ & \HeliNBinZeroPolariz &
\HeliNBinOnePolariz & \HeliNBinTwoPolariz \\ [0.5ex]
\hline
\hline
\end{tabular}
\label{tab:HelicityFitResults}
\end{center}
\end{table}

%
%
%
%
%

\ifthenelse{\boolean{confFlag}}{
\section{CONCLUSIONS}
}{
\section{Conclusions}
}
\label{sec:Conclusions}

Based on $\TotalBPairs\times 10^6$ $B\bar{B}$ pairs collected by the \babar\
detector at \pep2, we present a measurement of the $B^0\rightarrow
\bar{\Lambda} p \pi^-$ branching fraction and confirm the peaking of the
baryon-antibaryon mass spectrum near threshold, characteristic of
three-body baryonic B decays. In addition we measure the $\bar{\Lambda}$
polarization in \blppi\ decays as a function of \ELambdabar . The
measurement is compatible with the theoretical prediction
of full longitudinal right-handed polarization for large \ELambdabar,
which follows from the purely left-handed $b\rightarrow s$ transition in
the standard model \cite{ref:Suzuki_JPG29}.

\section{Acknowledgments}

We are grateful for the 
extraordinary contributions of our \pep2\ colleagues in
achieving the excellent luminosity and machine conditions
that have made this work possible.
The success of this project also relies critically on the 
expertise and dedication of the computing organizations that 
support \babar.
The collaborating institutions wish to thank 
SLAC for its support and the kind hospitality extended to them. 
This work is supported by the
US Department of Energy
and National Science Foundation, the
Natural Sciences and Engineering Research Council (Canada),
the Commissariat \`a l'Energie Atomique and
Institut National de Physique Nucl\'eaire et de Physique des Particules
(France), the
Bundesministerium f\"ur Bildung und Forschung and
Deutsche Forschungsgemeinschaft
(Germany), the
Istituto Nazionale di Fisica Nucleare (Italy),
the Foundation for Fundamental Research on Matter (The Netherlands),
the Research Council of Norway, the
Ministry of Education and Science of the Russian Federation, 
Ministerio de Educaci\'on y Ciencia (Spain), and the
Science and Technology Facilities Council (United Kingdom).
Individuals have received support from 
the Marie-Curie IEF program (European Union) and
the A. P. Sloan Foundation.

%
%
%
%
%
%
%

%
%
\ifthenelse{\boolean{authFlag}}{}{

\appendix
\newpage
\input{history_jour.tex}

}

\end{document}